\documentclass[conference]{IEEEtran}
\IEEEoverridecommandlockouts
\usepackage{cite}
\usepackage{amsmath,amssymb,amsfonts}
\usepackage{algorithmic}

\usepackage{multirow}
\usepackage{booktabs}
\usepackage[acronym]{glossaries}
\usepackage{textcomp}

\usepackage{graphicx}
\usepackage{subfigure}
\usepackage{xcolor}
\usepackage{amsmath}

\usepackage{bm}
\usepackage[colorlinks=false,pdfborder={0 0 0}]{hyperref}
\usepackage{float}
\usepackage{makecell}
\usepackage{nicefrac}
\usepackage{tikz,pgfplots}
\usepackage{tikzscale}
\usetikzlibrary{positioning, shapes, arrows}
\usepackage{stackengine}

\usepackage{cleveref}

\usepackage{array} % for "\extrarowheight" macro
\usepackage{pifont} % http://ctan.org/pkg/pifont, for "\cmark"
\usepackage{epsfig}

\newacronym{fvae}{FVAE}{factorized variational autoencoder}
\newacronym{gap}{GAP}{global average pooling}
\newacronym{cpc}{CPC}{contrastive predictive coding}
\newacronym{vtlp}{VTLP}{vocal tract length perturbation}
\newacronym{mse}{MSE}{mean squared error}
\newacronym{in}{IN}{instance normalization}
\newacronym{eer}{EER}{equal error rate}
\newacronym{air}{AIR}{Aachen Impulse Response}
\newacronym{ser}{SER}{Speech Emotion Recognition}
\newcommand{\spk}{\textrm{spk}}
\newcommand{\sty}{\textrm{sty}}
\usepackage{svg}

\begin{document}

\title{Speaker and Style Disentanglement of Speech Based on Contrastive Predictive Coding Supported Factorized Variational Autoencoder
}

\author{\IEEEauthorblockN{Yuying Xie\textsuperscript{1}, Michael Kuhlmann\textsuperscript{2}, Frederik Rautenberg\textsuperscript{2} , Zheng-Hua Tan\textsuperscript{1}, Reinhold Haeb-Umbach\textsuperscript{2}}
\IEEEauthorblockA{
    \textit{1. Department of Electronic Systems, Aalborg University, Denmark} \\
    \textit{2. Department of Communications Engineering, Paderborn University, Germany} \\
}
}

\maketitle

\begin{abstract}
\noindent
Speech signals encompass various information across multiple levels including content, speaker, and style. 
Disentanglement of these information, although challenging, is important for applications such as voice conversion.
The contrastive predictive coding supported factorized variational autoencoder achieves unsupervised disentanglement of a speech signal into speaker and content embeddings by assuming speaker info to be temporally more stable than content-induced variations.
However, this assumption may introduce other temporal stable information into the speaker embeddings, like environment or emotion, which we call style.
In this work, we propose a method to further disentangle non-content features into distinct speaker and style features, notably by leveraging readily accessible and well-defined speaker labels without the necessity for style labels.
Experimental results validate the proposed method's effectiveness on extracting disentangled features, thereby facilitating speaker, style, or combined speaker-style conversion. 
\end{abstract}


\section{Introduction}
Disentangled representation learning aims to extract features which can represent different attributes in the observed data. 
In speech signal processing, disentangled representation learning offers solutions extensively~\cite{ebbers2021contrastive, 9262021, du22c_interspeech, 9413512, sang2021deaan,li2022cross,qian2019autovc, qian2020unsupervised, randomcycle}, including voice conversion.
One essential property of speech is that the speaker's traits should remain consistent within a single utterance, while the content rapidly varies over time. This characteristic then can serve as a prior for developing an end-to-end voice conversion model to extract disentangled content and speaker features separately, as voice conversion aims to change the speaker trait and preserve the content of an utterance.

Ebbers et al.~\cite{ebbers2021contrastive} assume that the utterance-level information at the current frame is similar to that from the frames 1s before and after in the same utterance, and different between utterances in a batch. 
This assumption is then introduced as inductive bias in the \gls{fvae} to disentangle speaker from content information via a completely unsupervised fashion, requiring neither speaker labels nor a transcription. 
Specifically, during training, \gls{cpc} loss~\cite{oord2018representation} is applied on the utterance-level embeddings, by regarding feature of current frame as anchor, features from the same utterance as positive pairs and from other utterance in the batch as negative pairs.
Adversarial training is working on the content embedding for independence of extracted utterance-level and content features. 
The extracted utterance-level feature in \gls{fvae} are used to represent speaker identity only.

% sequential-level feature
Even though utterance-level feature is used for speaker identity representation widely~\cite{ECAPA-TDNN}, it can also contain other temporally stable information. 
For instance,  channel, emotion, or environment (e.g., room impulse response) information, should be similarly stable as speaker properties in one utterance.
Results in the works~\cite{xia2021self, cho22c_interspeech} about speaker embedding extraction also prove this point.
For instance, Xia et al.~\cite{xia2021self} utilize momentum contrastive (MoCo)~\cite{he2020momentum} learning on speaker embedding extraction. 
Results on the Voxceleb test set show that extracted features cluster not only on speaker identity, but also on session id.
Besides, Cho et al.~\cite{cho22c_interspeech} apply DIstillation with NO labels (DINO)~\cite{caron2021emerging} to extract utterance-level embeddings, and results show that the extracted embeddings can not only represent speakers, but also emotions.

Inspired by the fact that not only speaker traits change on the utterance-level, this work
further disentangles the utterance-level features from \gls{fvae}~\cite{ebbers2021contrastive} into speaker embedding and other temporally stable embeddings. 
We call the \textbf{other temporally stable} embeddings as \textbf{`style'} embeddings hereafter.  

The contribution of this work is as follows:
First, we posited that the speech signal is generated from three factors: speaker identity, style and content.
Second, this work extracts features representing these three factors in a hierarchical disentanglement manner.
The proposed method only needs speaker labels, which are cheap and easy to access. 
Third, the proposed method is applicable to different definitions of style embeddings.
In this work, acoustic environment and emotion have been chosen as different cases.
For the environment case, the performance of the proposed method is tested on both synthetic datasets and real unseen challenging datasets. 
For emotion, we evaluate the performance on a cross-language dataset.
Finally, we also investigated the application of the proposed method in style and speaker conversion using real-recorded datasets, both in clean and reverberant conditions.  \href{https://yuxi6842.github.io/speaker_style_disen.github.io/}{Demos showcasing these applications are provided.}\footnote{\href{https://yuxi6842.github.io/speaker_style_disen.github.io/}{https://yuxi6842.github.io/speaker\_style\_disen.github.io/}}

\section{Related work}

For fine-grained style control in speech synthesis and conversion, style  disentanglement is applied to obtain more expressive embeddings. Li et al.~\cite{li2022cross} propose a Tacotron2-based framework for cross-speaker and cross-emotion speech synthesis. 
Disentanglement is applied to get speaker-irrelavant and emotion-discriminative embeddings for emotion control. 
Although the model proposed in~\cite{li2022cross} can achieve cross-speaker emotion synthesis, only three speakers are contained in total.
Du et al.~\cite{du22c_interspeech} uses disentanglement to obtain speaker, content and emotion embeddings simultaneously for expressive voice conversion.
While the model in \cite{du22c_interspeech} uses mutual information loss for disentanglement, the evaluation is not sufficiently comprehensive, especially lacking in disentanglement analysis.
Besides, the extracted speaker features perform speaker verification poorly in~\cite{du22c_interspeech}.
Except for emotion, environment information like noise and reverberation has also been chosen as another controllable factor. 
Omran et al.~\cite{omran2023disentangling} shows an information bottleneck (IB) based way to split speech and noise or reverberation. Masking is used on specific dimensions in the latent variable space, while reconstruction loss controls the particular information flow.
However, similar to other IB based methods, the performance depends heavily on bottleneck design in~\cite{omran2023disentangling}. 
Another work in~\cite{zhang2022metaspeech} uses disentanglement to realize environment conversion, in which only four seen environments are considered in their work.

% %%%%%%%%%%%%%%%%%%%%%%%%%%%%%%%%%%%%%%%%%%%%%%%%%%%%%%%%%%%%%%%%%%%%%%%%%%%%%%
% %                             Proposed method                                %
% %%%%%%%%%%%%%%%%%%%%%%%%%%%%%%%%%%%%%%%%%%%%%%%%%%%%%%%%%%%%%%%%%%%%%%%%%%%%%%

\section{Proposed method}

\subsection{Theoretical Explanation}

Disentanglement aims to find codes to represent underlying causal factors~\cite{bengio2013representation}.
The original \gls{fvae} actually assumes the log-mel feature $\mathbf{X}=[\mathbf{x}_1, \ldots, \mathbf{x}_T]$ is generated from independent latent variables $\mathbf{S}$ and $\mathbf{Z}$ via a random generative process:
\begin{equation}
    \label{eq:generation of FVAE} \mathbf{X}=g_1(\mathbf{S},\mathbf{Z}),
\end{equation}
in which $\mathbf{S}$ and  $\mathbf{Z} $ represent the utterance-level characteristic and content, respectively.

\begin{figure}[htbp]
\vspace{-1em}
    \centering
    \subfigure[inference process]{
    \label{figx:a}
    \includegraphics[height=52pt]{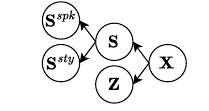}
    }
    \subfigure[generation process]{
    \label{figx:b}
    \includegraphics[height=60pt]{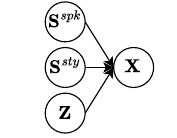}
    }
    \caption{Graphical illustration of the proposed method. The inference process follows a hierarchical structure.} 
  \label{fig:graphical illustration}
\end{figure}

In this work, we further assume latent variable $\mathbf{S}$ is generated from a random process with
\begin{equation}
\label{eq: generation of s}
 \mathbf{S}=g_2(\mathbf{S}^{\textrm{spk}}, \mathbf{S}^{\textrm{sty}}),   
\end{equation}
in which $\mathbf{S}^{\textrm{spk}}$ and $\mathbf{S}^{\textrm{sty}}$ are independent latent variables to denote speaker identity and style in one utterance.
Thus eq~\eqref{eq:generation of FVAE} can also be written as:
\begin{equation}
\label{eq:generation of x with S} 
\mathbf{X}=g_1(\mathbf{S},\mathbf{Z})=g_1(g_2(\mathbf{S}^{\textrm{spk}}, \mathbf{S}^{\textrm{sty}}),\mathbf{Z}),
\end{equation}
and can be further simplified as:
\begin{equation}
\label{eq:generation of x with all} 
\mathbf{X}=g(\mathbf{S}^{\textrm{spk}}, \mathbf{S}^{\textrm{sty}},\mathbf{Z}).
\end{equation}

Generation process of the proposed method illustrated in fig~\ref{figx:b}  is learned as eq.~\eqref{eq:generation of x with all}.
Meanwhile, inference process in the proposed method is based on eq~\eqref{eq:generation of x with S} following a hierarchical structure as in fig~\ref{figx:a}.
The first step, inherited from \gls{fvae}, learns to decompose observation $\mathbf{X}$ into latent variables $\mathbf{S}$ and $\mathbf{Z}$ according to the data inherent properties.
The second step further disentangles $\mathbf{S}$ into $\mathbf{S}^{\textrm{sty}}$ and $\mathbf{S}^{\textrm{spk}}$ by introducing low-cost speaker labels in training. 
Adversarial learning is applied in both steps to achieve independence.
As self-supervised learning method have been used in the first disentanglement step, the proposed hierarchical structure reduces the labeling cost of disentangling multiple factors.

\subsection{Structure}
\label{sec:proposed method}
\begin{figure}[t]
  \centering
  \includegraphics[width=1.1\columnwidth]{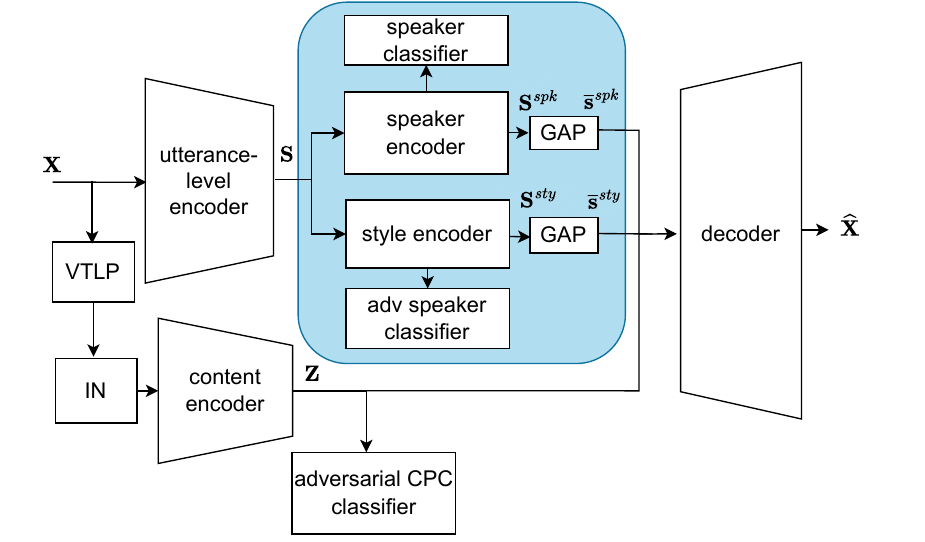}
  \caption{Diagram of the proposed method. Highlighted in blue is the proposed disentanglement in speaker and style properties. VTLP stands for vocal tract length perturbation, IN for instance normalization, CPC for contrastive predictive
coding, and GAP for global average pooling.}
  \label{fig:structure of proposed method}
  \vspace{-1em}
\end{figure}

The structure of the proposed method is shown in~\cref{fig:structure of proposed method}. 
Compared with FVAE, the new component, the proposed disentanglement module is highlighted in blue in~\cref{fig:structure of proposed method}.
The proposed module contains a speaker encoder, a speaker classifier, a style encoder and an adversarial speaker classifier.
Additionally, like FVAE, both \gls{vtlp}~\cite{jaitly2013vocal} and \gls{in}~\cite{ulyanov2016instance} are used to partially remove speaker information at the content encoder's input.
The utterance-level encoder, the content encoder, the adversarial CPC classifier, and the decoder are same with ~\cite{ebbers2021contrastive}. 

The first disentanglement step, inherited from FVAE, is applied on $\mathbf{X}$ to get utterance-level feature $\mathbf{S}=[\mathbf{s}_1, \ldots, \mathbf{s}_T]$ and content feature $\mathbf{Z}=[\mathbf{z}_1, \ldots, \mathbf{z}_N]$. 
The contrastive learning method is firstly applied to $\mathbf{S}$, to ensure that the extracted utterance-level feature is temporally stable in one utterance.
Regarding embedding $\mathbf{s}_t$ as an anchor, $\mathbf{s}_{t+\tau}$ from the same utterance as the positive pair, and $\tilde{\mathbf{s}}_{t+\tau}$ from other utterances in the training batch $\mathcal{B}$ as the negative pair, \gls{cpc} loss is then calculated on $\mathbf{S}$ as:
\begin{equation}
	L_{\mathrm{cpc}} = -\frac{1}{T - \tau } \sum_{t=1}^{T-\tau} \frac{\mathrm{exp}({\mathbf{s}_{t+\tau}^\mathrm{T} \cdot  \mathbf{s}_{t}})}{\sum_{\mathcal{B}} \mathrm{exp}({\tilde{\mathbf{s}}_{t+\tau}^\mathrm{T} \cdot \mathbf{s}_{t}})} \,
 \label{cpc}
\end{equation}
in which $T$ is the frame length of input sequence $\mathbf{X}$, and $\tau=80$ (corresponding to 1s) is the time lag.
Variational autoencoder (VAE)~\cite{vae} is applied on modeling content embedding $\mathbf{Z}$ as an information bottleneck approach for improving disentanglement.
The length of extracted $\mathbf{Z}$ is $N$. $N\leq T$ as there may be downsampling in content encoder.
Each vector $\mathbf{z}_n$ in $\mathbf{Z}$ is regarded as a stochastic variable with prior $p(\mathbf{z}_n)=\mathcal{N}(\mathbf{0}, \mathbf{I})$ and posterior $q(\mathbf{z}_n)=\mathcal{N}(\bm{\mu}_n, \textrm{diag}(\bm{\sigma}_n^2))$. Parameters of $q(\mathbf{z}_n)$ are generated from the content encoder and optimized with KL divergence:
\begin{equation}
    L_{\mathrm{kld}} = \frac{1}{N} \sum_{n=1}^{N} \mathrm{KL}(q(\mathbf{z}_n)\|p(\mathbf{z}_n)).
    \label{kld}
\end{equation}
Besides, an adversarial CPC classifier is applied on $\mathbf{Z}$ for independence of $\mathbf{S}$, cf. ~\cite{ebbers2021contrastive}.

The second disentanglement is then operated on the frame-wise utterance-level features $\mathbf{S}$ via the proposed new module.
Speaker and style encoders have same structure built on one-dimensional CNN layers.
Denote
${\mathbf{S}^{\spk}=[\mathbf{s}^{\spk}_1, \ldots, \mathbf{s}^{\spk}_T]}$ and 
$\mathbf{S}^{\sty}=[\mathbf{s}^{\sty}_1, \ldots, \mathbf{s}^{\sty}_T]$ as the output from speaker encoder and style encoder.
Speaker classification then works frame-wisely on $\mathbf{s}^{\spk}_t$ to encourage extracting speaker information alone.
Meanwhile, frame-wise adversarial speaker classification is applied on $\mathbf{s}^{\sty}_t$ to ensure that the extracted style embedding is less dependent on speaker trait.
Gradient reversal layer~\cite{GRL} is applied between the output layer of the style encoder and the adversarial speaker classifier. 
Cross-entropy loss is used here for classification on both $\mathbf{s}^{\spk}_t$ and $\mathbf{s}^{\sty}_t$.
To get aggregated information over time and discard unnecessary details, vectors $\overline{\mathbf{s}}^{\spk}$ and $\overline{\mathbf{s}}^{\sty}$ are obtained by applying \gls{gap} on $\mathbf{S}^{\spk}$ and $\mathbf{S}^{\sty}$.

The decoder takes as input a matrix where $\overline{\mathbf{s}}^{\spk}$ and $\overline{\mathbf{s}}^{\sty}$ are repeated $N$ times along the time axis and then concatenated with the content embedding $\mathbf{Z}$.
To get sharper, less-oversmoothed output, reconstruction loss here is XSigmoid function~\cite{beck2022wavebender}:
\begin{equation}
\label{eq: XSigmoid function}
    L_{\mathrm{rec}} = \frac1T\sum_t\Vert|\mathrm{\hat{\mathbf{x}}}_t - \mathrm{\mathbf{x}}_t| \left(2 \sigma(\mathrm{\hat{\mathbf{x}}}_t - \mathrm{\mathbf{x}}_t\right) - 1)\Vert_1\, ,
\end{equation}
in which $\sigma(\cdot)$ is the sigmoid function.

Denote $\boldsymbol{\theta}_{\mathrm{enc}}^{\mathrm{utt}}$, $\boldsymbol{\theta}_{\mathrm{enc}}^{\mathrm{spk}}$, $\boldsymbol{\theta}_{\mathrm{enc}}^{\mathrm{sty}}$, $\boldsymbol{\theta}_{\mathrm{enc}}^{\mathrm{cont}}$ are the parameter sets of utterance encoder, speaker encoder, style encoder and content encoder.
And the union of all encoders' parameter sets as
$\boldsymbol{\theta}_{\mathrm{enc}}=\boldsymbol{\theta}_{\mathrm{enc}}^{\mathrm{utt}} \cup \boldsymbol{\theta}_{\mathrm{enc}}^{\mathrm{spk}} \cup \boldsymbol{\theta}_{\mathrm{enc}}^{\mathrm{sty}} \cup \boldsymbol{\theta}_{\mathrm{enc}}^{\mathrm{cont}}$. 
The parameter sets of speaker classifier and adversarial speaker classifier are $\boldsymbol{\theta}_{\mathrm{clf}}^{\mathrm{spk}}$, $\boldsymbol{\theta}_{\mathrm{adv\_clf}}^{\mathrm{spk}}$, while the parameter set of decoder is $\boldsymbol{\theta}_{\mathrm{dec}}$
The total loss function is then written as:
\begin{equation}
\label{eq: proposed loss}
\begin{aligned}
        &L(\boldsymbol{\theta}_{\mathrm{enc}},
        \boldsymbol{\theta}_{\mathrm{dec}},
        \boldsymbol{\theta}^{\mathrm{spk}}_{\mathrm{clf}}, \boldsymbol{\theta}^{\mathrm{spk}}_{\mathrm{adv\_clf}})  \\
        & = L_\mathrm{rec}(g_{\hat{\mathbf{X}}}(\mathbf{X}; \boldsymbol{\theta}_{\mathrm{enc}},\boldsymbol{\theta}_{\mathrm{dec}}),\mathbf{X}) 
        + \lambda_s L_{\mathrm{cpc}}(g_{\mathbf{S}}(\mathbf{X}; \boldsymbol{\theta}_{\mathrm{enc}}^{\mathrm{utt}}))\\
        & + \beta L_{\mathrm{kld}}(g_{\mathbf{Z}}(\mathbf{X};\boldsymbol{\theta}_{\mathrm{enc}}^{\mathrm{cont}}))
         -  \lambda_z L_{\mathrm{cpc}} (R(g_{\mathbf{Z}}(\mathbf{X}; \boldsymbol{\theta}_{\mathrm{enc}}^{\mathrm{cont}})))
        \\
        &+ L_{\mathrm{CE}}(g_{\mathbf{P}^{\mathrm{spk}}
        }(g_{\mathbf{S}^{\mathrm{spk}}}(\mathbf{X};\boldsymbol{\theta}_{\mathrm{enc}}^{\mathrm{utt}},\boldsymbol{\theta}_{\mathrm{enc}}^{\mathrm{spk}});\boldsymbol{\theta}_{\mathrm{clf}}^{\mathrm{spk}}),\mathbf{Y})
        \\
        & - L_{\mathrm{CE}}(g_{\mathbf{P}^{\mathrm{sty}}}(R(g_{\mathbf{S}^{\mathrm{sty}}}(\mathbf{X}; \boldsymbol{\theta}_{\mathrm{enc}}^{\mathrm{utt}},\boldsymbol{\theta}_{\mathrm{enc}}^{\mathrm{sty}}));\boldsymbol{\theta}_{\mathrm{adv\_clf}}^{\mathrm{spk}}),\mathbf{Y}),
\end{aligned}
\end{equation}
in which $\mathbf{Y}$ is the true speaker label one-hot matrix, $\mathbf{P}^{\mathrm{spk}}$ and $\mathbf{P}^{\mathrm{sty}}$ are speaker label predictions from speaker classifier and adversarial speaker classifier. 
Mapping $g_{\mathbf{y}}(\mathbf{x};\boldsymbol{\theta})$ denotes that $\mathbf{y} = g_{\mathbf{y}}(\mathbf{x};\boldsymbol{\theta})$, where $\mathbf{x}$ is the input and $\boldsymbol{\theta}$ is the parameter set in the corresponding neural networks.
For a better illustration of forward and backward behavior,
gradient reversal layer is represented via a 'pseudo-function' $R(\cdot)$ as~\cite{GRL_math}:
\begin{equation}
    R(\mathbf{x}) = \mathbf{x}, \frac{\mathrm{d}R(\mathbf{x})}{\mathrm{d}\mathbf{x}} = - \mathbf{I},
\end{equation}
where $\mathbf{I}$ is an identity matrix. 
To balance the terms, $\lambda_s$, $\lambda_z$ and $\beta$ are coeffients in the total loss function.

%%%%%%%%%%%%%%%%%%%%%%%%%%%%%%%%%%%%%%%%%%%%%%%%%%%%%%%%%%%%%%%%%%%%%%%%%%%%%%
%                             Experiment                                     %
%%%%%%%%%%%%%%%%%%%%%%%%%%%%%%%%%%%%%%%%%%%%%%%%%%%%%%%%%%%%%%%%%%%%%%%%%%%%%%
\section{Experiments}
\label{sec:experiment}
% dataset
\subsection{Datasets}
\label{exp: dataset}
% librispeech (train + voice convesion)

For testing the separation of speaker from environment information, the model is trained and tested with artificially reverberated data. 
Besides, an unseen real-recorded dataset is chosen for testing only.
For the case of separating speaker from emotion information, a cross-language emotion dataset is used for training and testing.

\noindent
\textbf{LibriSpeech}: 
The clean speech dataset of this work is based on LibriSpeech~\cite{7178964}.
The subsets \textit{train-clean-360} (921 speakers), \textit{train-clean-100} (251 speakers) and \textit{test-clean} (40 speakers) are used and convolved with real-recorded RIRs.

\noindent \textbf{LibriSpeech + \Gls{air}}:
The \Gls{air} dataset~\cite{5201259} contains 107 real RIR recordings, and is used for data augumentation on the training dataset.
Each utterance from the LibriSpeech training set is convolved with 4 different randomly selected RIRs from \gls{air}. 
Thus, the training dataset in the environment case is 4 times larger than the LibriSpeech training set.
% RT60

\noindent \textbf{VOiCES}: 
To evaluate the performance on speaker verification, 
VOiCES~\cite{richey2018voices} is chosen in this work \textit{for testing only}. 
This dataset is recorded in four different rooms, with differently-located microphones and various directions of arrival between microphones and speakers.
In this work, only the speech dataset without noise from \textit{VOiCES\_devkit} is used for evaluation on real challenging acoustic environment. 

\noindent \textbf{Emotion dataset}: The emotional speech dataset (ESD)~\cite{zhou2021seen} contains studio-quality recordings of 10 English and 10 Mandarin speakers in neutral style and four acted emotions (happy, angry, sad, and surprise).
Each speaker utters the same 350 utterances in all five styles, totalling 35,000 utterances.
The subset which contains all 10 English speakers and 4 styles (neutral, happy, angry, and sad) is used for training, validation and evaluation (seen condition) firstly, with split of 85\%/10\%/5\%, respectively.
Besides, the remaining style, surprise, is used for evaluation as an unseen emotion. Moreover, the Mandarin subset is used for evaluation when both speaker and language are unseen.

\begin{table*}[t]
    \centering
    \caption{%
        Speaker verification and speech emotion recognition (Emo Rec) results on ESD.
        We measure the EER when same speaker pairs are compared with the same emotion (within-emotion, WE) or with different emotion (across-emotion, AE).
    }
    \begin{tabular*}{\textwidth}{@{\extracolsep{\fill}}l c  cc cc cc cc}
        \toprule[1.5pt]
        \multirow{3}*{\textbf{Framework}} & \multirow{3}*{\textbf{Embedding}} & \multicolumn{6}{c}{\textbf{Speaker Verification Equal Error Rate (EER)}(\%)~$\downarrow$  } & \multicolumn{2}{c}{\textbf{Emo Rec Acc.}(\%)~$\uparrow$} \\
        \cmidrule[1pt]{3-8} \cmidrule[1pt]{9-10}
        & & \multicolumn{2}{c}{Engl., w/o. surprise} & \multicolumn{2}{c}{Engl., all} & \multicolumn{2}{c}{Mand., all} & \multirow{2}*{Engl.} & \multirow{2}*{Mand.} \\
        \cmidrule[1pt]{3-4} \cmidrule[1pt]{5-6} \cmidrule[1pt]{7-8}
        & &  WE & AE & WE & AE & WE & AE & & \\
        \midrule[1pt]
        FVAE & utterance-level &  0.14 & 0.27 & 0.28 & 0.63 & \textbf{12.69} & 31.89 & 89.43 & \textbf{31.36} \\
        \midrule
        % \multirow{2}*{proposed} & speaker &   \textbf{0.00} & \textbf{0.00} & \textbf{0.17} & \textbf{0.63} & 16.77 & 32.73 & 86.70 & 24.77 \\
        \multirow{2}*{Proposed} & speaker & \textbf{0.00} & \textbf{0.00} & \textbf{0.21} & \textbf{0.44} & 14.65 & \textbf{26.35} & 91.36 & 23.64 \\
        % & style &  30.49 & 55.20 & 31.09 & 52.20 & 24.29 & 48.49 & \textbf{89.77} & \textbf{36.59} \\
        & style &  1.82 & 11.62 & 2.24 & 11.86 & 16.10 & 41.84 & \textbf{92.27} & 30.57 \\
        % \midrule
        % FVAE & style & LibriSpeech & \cmark & 5.13\% & 15.14\% & \textbf{5.47\%} & \textbf{13.69\%} & 5.44\% & \textbf{24.86\%} \\
        % \multirow{2}*{proposed} & speaker & \multirow{2}*{LibriSpeech} & \multirow{2}*{\cmark} & \textbf{4.02\%} & \textbf{12.33\%} & 5.70\% & 15.98\% & \textbf{5.19\%} & 27.33\% \\
        % & style & & & 27.14\% & 53.19\% & 28.38\% & 51.03\% & 18.78\% & 45.52\% \\
        \bottomrule[1.5pt]
    \end{tabular*}
    \label{tab:esd-sv}
\end{table*}

\subsection{Implementation Details}
Following~\cite{ebbers2021contrastive}, each update of the encoders, speaker classifier and decoder is followed by three exclusive updates of the adversarial CPC and adversarial speaker classifier.
We set $\beta=0.01$ and $\lambda_s=\lambda_z=1$.
We use the Adam optimizer and a learning rate of $5\times10^{-4}$. Both \gls{fvae} and proposed method use eq~\eqref{eq: XSigmoid function} as reconstruction loss for fair.

\noindent \textbf{Environment}
Training is based on the 'LibriSpeech + AIR' set with eq~\eqref{eq: proposed loss} as loss function. 
The utterance-level encoder, content encoder and decoder in this work have the same structure as in~\cite{ebbers2021contrastive}. 
The speaker encoder and style encoder both contain 3 one-dimensional CNN layers with stride equal to 1 and kernel size equal to 5 for all layers. The extracted speaker embedding and style embedding have both a dimension of 128. The speaker classifier contains one fully-connected (FC) layer, while the adversarial speaker classifier contains three FC layers with 128 hidden units each. The other settings are the same as in~\cite{ebbers2021contrastive}.

\noindent \textbf{Emotion}
Training on ESD is performed for 100,000 iterations.
The networks are configured in the same way as in environment case, except for the input of content encoder: Inspired by HuBERT~\cite{hsu2021hubert}, features from a CNN-based waveform extractor are used to replace the mel spectrogram.
The whole model is firstly pre-trained on LibriTTS~\cite{zen2019libritts}. 
Then fixed the content encoder, and fine-tuned the other modules in the whole model on ESD. 
The loss functions for pretraining and fine-tuning are the same, i.e. eq~\eqref{eq: proposed loss}.

\setlength{\tabcolsep}{2mm}
\begin{table}[t]
  \caption{Speaker verification results under environment case. For LibriSpeech dataset, EER is calculated when same speaker pairs are selected within the same chapter (WC) and across chapter (AC).}
  \label{tab: environment sv}
  \centering
  %\begin{tabular}{*{4}{>{\centering\arraybackslash}X}}
  \begin{tabular}{ c| c | c| c}
    \toprule[1.5pt]
    \multicolumn{2}{c}{\textbf{Framework}} & \multicolumn{2}{|c}{\textbf{EER} (\%)~$\downarrow$} \\
    \cline{3-4} 
    \multicolumn{2}{c|}{ } & WC & AC\\
    \cline{1-4}
    \multirow{3}{*}{LibriSpeech} & FVAE & $2.44$ & $6.09$ ~~~ \\ \cline{2-4}
    & proposed(before disen) & $2.96$ & $6.82$ ~~~ \\ \cline{2-4}
    & proposed (speaker) & $\bm{1.50}$ & $\bm{4.65}$ ~~~ \\ \cline{2-4}
    & proposed (style) & $18.65$ & $24.29$ ~~~ \\ \cline{1-4}
    \multirow{3}{*}{\makecell{VOiCES \\ (test only)} } & proposed(before disen) & \multicolumn{2}{c}{$20.6$} ~~~ \\ \cline{2-4}
    & proposed (speaker) & \multicolumn{2}{c}{\bm{$12.5$}} ~~~ \\ \cline{2-4}
    & proposed (style) & \multicolumn{2}{c}{$25.7$} ~~~ \\  
    \bottomrule[1.5pt]
  \end{tabular}
  \vspace{-1em}
\end{table}

% embedding disentanglement

\subsection{Evaluation under Environment Case}
\begin{figure}[htbp]
\vspace{-1em}
    \centering
    \subfigure[style embeddings with colors denoting RIR labels]{
    \label{fig1:a}
    \includegraphics[width=0.4\columnwidth]{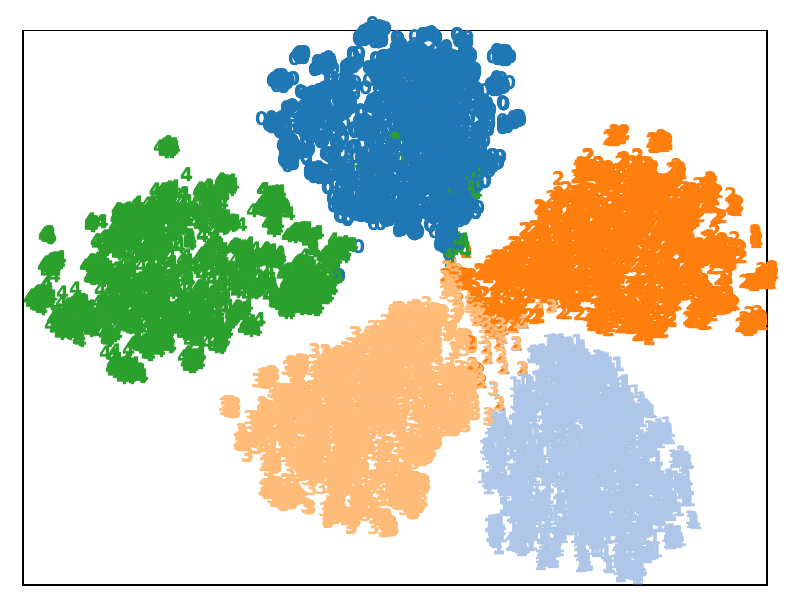}
    }
    \subfigure[speaker embeddings with colors denoting RIR labels]{
    \label{fig1:b}
    \includegraphics[width=0.4\columnwidth]{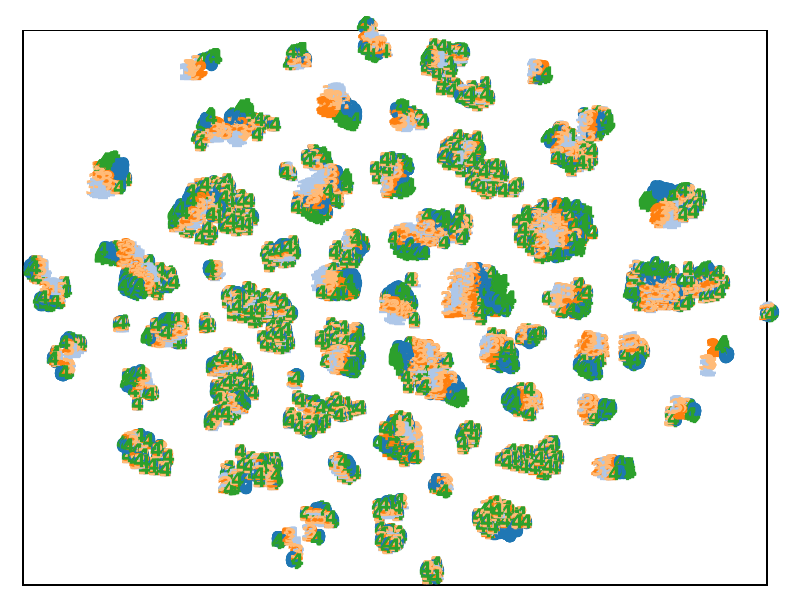}
    } \\
    \subfigure[style embeddings with colors denoting speaker labels]{
    \label{fig1:c}
    \includegraphics[width=0.4\columnwidth]{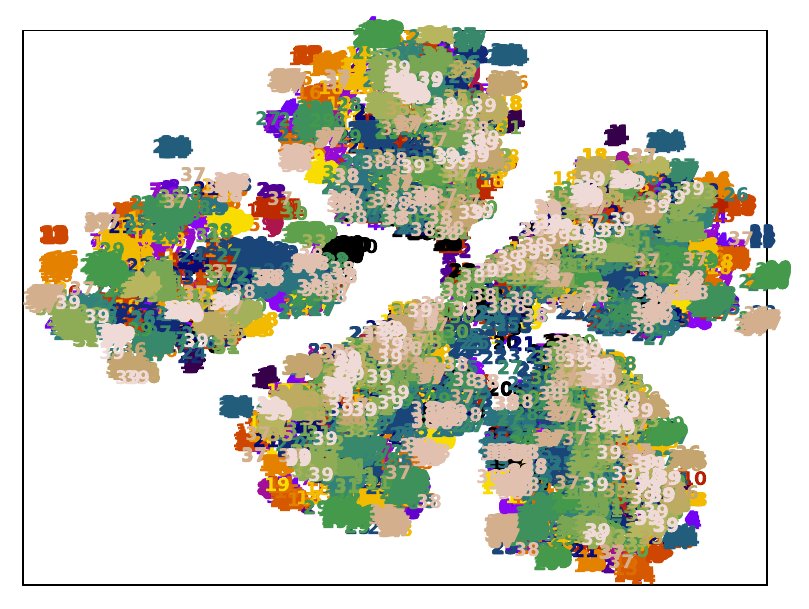}
    }
    \subfigure[speaker embeddings with colors denoting speaker labels]{
    \label{fig1:d}
    \includegraphics[width=0.4\columnwidth]{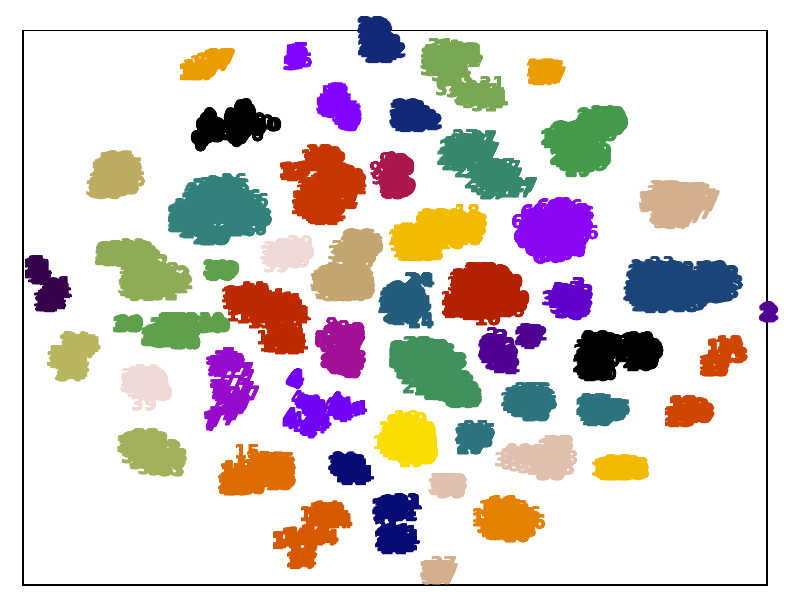}
    }
    \caption{t-SNE plots of embeddings under environment case. One can clearly see that style embeddings cluster according to RIR labels and speaker embeddings according to speaker labels.}
    \label{fig:t-SNE}
    \vspace{-1em}
\end{figure}

We firstly use t-SNE plots to visualize the extracted style and speaker embeddings.
Results are shown in fig~\ref{fig:t-SNE}.
The example shown is taken from the LibriSpeech \textit{test-clean} set, convolved with 4 different RIRs from the AIR dataset. 
These figures show that the style and speaker features form well-defined clusters according to their  style and speaker labels, respectively, while speakers are spread across style clusters, and RIR-labels across speaker clusters.
This demonstrates that the proposed second disentanglement can separate speaker and environment information well.

% Comment to increase compilation seed
% Uncomment to include in final version

Moreover, we test the extracted embeddings performance via speaker verification tests on the LibriSpeech \textit{test-clean} set and VOiCES \textit{dev-kit} dataset. 
Evaluation is based on equal error rate (EER) according to cosine similarity. 
Results are shown in~\cref{tab: environment sv}.
When dealing with Librispeech test set, we found that there is slight recording environment change between some chapters in the original dataset.
Thus, EER is calculated separately for the case when same speaker pairs are selected within the same chapter (WC) and when selected across chapters (AC), with the latter giving larger EERs.
Compared with original \gls{fvae}, the proposed speaker embeddings after the second disentanglement step result in an improvement in EER by a margin on the clean LibriSpeech dataset, cf.~\cref{tab: environment sv}. 
Even though the proposed speaker embeddings reduce the gap between WC and AC, the exist of the gap may because: compared to the generated data, environmental changes in LibriSpeech can be negligible.  

Besides, the proposed method shows effective disentanglement on VOiCES \textit{dev-kit}
dataset. The speaker embedding from proposed second disentanglement shows large improvement with an EER of $12.5\%$ compared with $20.6\%$ before the second disentanglement. 
Even though VOiCES is unseen during training, this result illustrates that the second disentanglement can help on extracting more robust speaker embeddings under challenging realistic situation.

\subsection{Evaluation under Emotion Case}
Both speaker verification and speech emotion recognition results are shown in~\cref{tab:esd-sv}.
To check how the extracted features influenced by emotion, we calculate EER in speaker verification experiment under two different scenarios when positive speaker utterance pairs have the same emotion (within-emotion, WE) or not (across-emotion, AE).
For emotion recognition, we train a 3-layer FC classifier on the embeddings using the English speaker subset with all 5 emotional states.

In speaker verification experiments, when speaker and language are seen in training, speaker embedding performs superior compared with utterance-level features and style embedding.
Meanwhile, the evaluation of unseen speakers and language (Mandarin) subset, reveals that speaker embedding demonstrates stability and is less prone to the impact of emotional factors in speech. This is especially evident when positive pairs in speaker verification involve varying emotions.
For speech emotion recognition, style embedding shows slightly better performance when language and speaker are seen during training, and comparable performance with utterance-level features in Mandarin subset.

%%%%%%%%%%%%%%%%%%%%%%%%%%%%%%%%%%%%%%%%%%%%%%%%%%%%%%%%%%%%%%%%%%%%%%%%%%%%%%
%                             Conclusion                                     %
%%%%%%%%%%%%%%%%%%%%%%%%%%%%%%%%%%%%%%%%%%%%%%%%%%%%%%%%%%%%%%%%%%%%%%%%%%%%%%

\section{Conclusion}
\label{sec:conclusion}
% 1. contribution
% 2. experiment results
Inspired by the existing results that the utterance-level feature contains not only speaker identity, but other style attributes, this work proposed a hierarchical disentanglement method based on \gls{fvae}.
Particularly, the utterance-level embedding from \gls{fvae} is further decomposed into speaker and style features in this work.
Training of this disentanglement framework  needs speaker labels only, which are easy to get.
We evaluated the proposed method under two cases of style: environment and emotion. 
For the environment case, the extracted style embeddings show clusters on RIR labels.
The extracted speaker embeddings perform well for  speaker verification, when tested on a clean speech dataset and a real-recorded reverberation dataset, indicating the effectiveness of the proposed method.
The findings from the emotion dataset demonstrate that the extracted speaker embeddings become more distinctive in terms of speaker identification and less susceptible to emotional variations. 
We hope this work may contribute to future work including multi-factor voice conversion, and conversion under challenging environments.

\bibliographystyle{IEEEbib}
\bibliography{refs}

\end{document}